\documentclass[aps,prl,groupedaddress,superscriptaddress,twocolumn]{revtex4-1}
\usepackage{graphicx}
\usepackage{amssymb}

\begin{document}

\title{Nanopattern-stimulated superconductor-insulator transition in thin TiN films}

\author{T.\,I. Baturina}
\affiliation{ A. V. Rzhanov Institute of Semiconductor Physics SB RAS, 13 Lavrentjev Avenue, 
Novosibirsk, 630090 Russia}
\affiliation{Novosibirsk State University, 2 Pirogova Street, Novosibirsk, 630090 Russia}
\affiliation{Materials Science Division, Argonne National Laboratory, Argonne, Illinois 60439, USA}
\author{V.\,M. Vinokur}
\affiliation{Materials Science Division, Argonne National Laboratory, Argonne, Illinois 60439, USA}
\author{A.\,Yu. Mironov}
\affiliation{ A. V. Rzhanov Institute of Semiconductor Physics SB RAS, 13 Lavrentjev Avenue, 
Novosibirsk, 630090 Russia}
\author{N.\,M.~Chtchelkatchev}
\affiliation{Materials Science Division, Argonne National Laboratory, Argonne, Illinois 60439, USA}
\affiliation{Institute for High Pressure Physics, Russian Academy of Sciences, 
Troitsk 142190, Moscow region, Russia}
\author{D.\,A.~Nasimov}
\affiliation{ A. V. Rzhanov Institute of Semiconductor Physics SB RAS, 13 Lavrentjev Avenue, 
Novosibirsk, 630090 Russia}
\author{A.\,V.~Latyshev}
\affiliation{ A. V. Rzhanov Institute of Semiconductor Physics SB RAS, 13 Lavrentjev Avenue, 
Novosibirsk, 630090 Russia}
\affiliation{Novosibirsk State University, 2 Pirogova Street, Novosibirsk, 630090 Russia}

\date{\today}

\begin{abstract}
We present the results of the comparative study of the influence of disorder 
on transport properties in continuous and nanoperforated TiN films.  
We show that nanopatterning turns a thin TiN film into an array 
of superconducting weak links and stimulates both, the disorder- and magnetic 
field-driven superconductor-to-insulator transitions, pushing them to lower degree of disorder.  
We find that nanopatterning enhances the role of the two-dimensional Coulomb interaction 
in the system transforming the originally insulating film into a more pronounced insulator.
We observe magnetoresistance oscillations reflecting collective behaviour 
of the multiconnected nanopatterned superconducting 
film in the wide range of temperatures and uncover the physical mechanism 
of these oscillations as phase slips in superconducting weak link network.
\end{abstract}

\pacs{74.78.-w, 74.81.Fa, 74.81.-g, 74.62.En, 74.40.Kb}


\maketitle
That a thin film of the same material can be a superconductor but can very well 
turn an insulator, is one of the most remarkable aspects of disordered 
superconductors~\cite{Strongin70,Haviland,Goldman1993,GoldmanReview,BeWu,TiNSanquer,TaYoon}.  
The engine driving the transition between the superconducting and insulating states  
is disorder the effect of which is two-fold.
On the one hand, disorder limits the electron diffusion enhancing 
thus the Coulomb electron-electron interaction which competes with 
the Cooper pairing~\cite{Maekawa,Finkelstein}.
The latter in an interplay with the disorder-induced inhomogeneities 
localizes Cooper pairs to form an insulating state, Cooper-pair insulator.  
A restricted geometry is critical to effects of disorder -- 
for the insulating state to be observed the superconducting material 
is to be thinned down till its thickness $d$ becomes comparable 
to  or smaller than the superconducting coherence length $\xi$.
One of the major experimental challenges in these studies remains 
the optimization of material parameters taking it to the closest proximity of 
the direct  superconductor-insulator transition and identifying the systems
that exhibit such a transition at available temperatures.  In this Letter we 
meet this challenge via creating a metamaterial with the desirable properties,
the multiconnected thin superconducting film.
We show that nanopatterning a thin TiN film into a regular 
sieve-like configuration turns it into an array of weak links and, therefore,
stimulates the direct superconductor-to-insulator transition. 
Depending on the original degree of disorder it either suppresses 
the critical temperature $T_c$, or drives the initially superconducting film 
into an insulating state, or else,  transforms the originally insulating film 
into an even more pronounced insulator.

\begin{figure}[b]
\includegraphics[width=1.0\columnwidth]{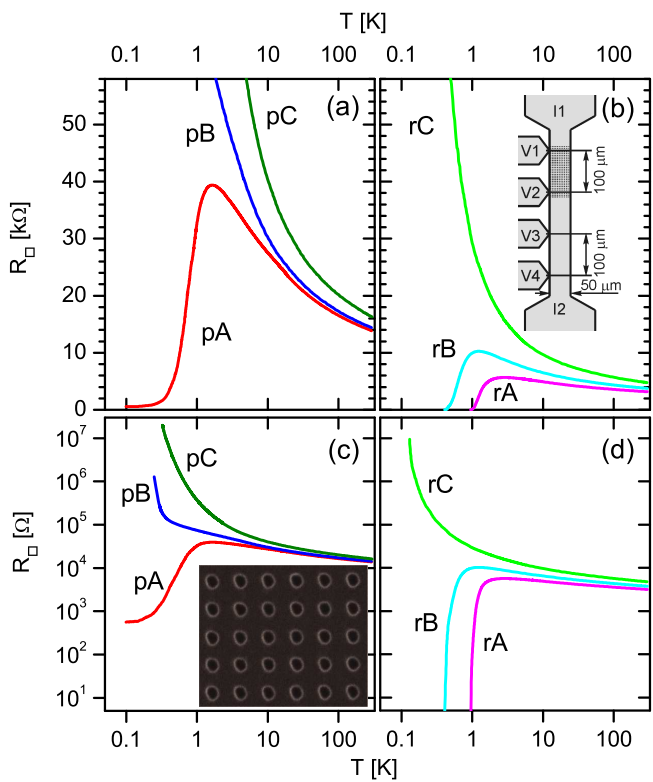}
\caption{\label{fig:RT} 
Resistance per square versus temperature plots 
displaying superconductor-insulator transitions in the perforated films [panels (a) and (c)] 
and the corresponding reference films [panels (b) and (d)]. 
Panels (a) and (b) employ the linear scales of the resistance, 
whereas in the panels (c) and (d) the logarithmic scales are used.  
The scales of the resistances are the same pairwise for the (a)-(b) and (c)-(d) panels.
The inset to panel (b) sketches the measuring set up (the dimensions shown):
I1 and I2 are the current-carrying and the pick up leads, respectively, 
V1 and V2 leads are attached within the perforated part of the film and 
V3 and V4 pick up voltage from the original non-perforated section of the film. 
The inset to panel (c) presents scanning electron microscope (SEM) 
image of the part of the TiN film patterned with a square array of holes with the 
diameter $\sim 120$\,nm and with the center spacing of 200\,nm.
}
\end{figure}

As a starting material we have chosen a 5\,nm thin TiN film which was identical by its parameters 
to those that experienced the superconductor-insulator transition after soft plasma 
etching~\cite{TiNPhysB2005,QMTiNPRL,SITTiNPRL,QRCTiNPhysB} 
and which were fully characterized by the high resolution electron beam, 
infrared~\cite{OpticsTiN}, 
and low-temperature scanning tunnelling spectroscopy~\cite{STM_TiN}.
The smooth, continuous, and uniform TiN film was
formed on the Si/SiO$_2$ substrate by atomic layer deposition.
The film had the superconducting critical temperature $T_{\mathrm c}=1.03$\,K, 
the diffusion constant $D=0.3$\,cm$^2$/s, 
and the superconducting coherence length $\xi_{\mathrm d}(0)=9.3$\,nm.
As a first step, the film was patterned into the bridges 50\,$\mu$m wide 
and 100\,$\mu$m long via conventional UV lithography.  
Then, making use of the electron lithography and the subsequent plasma etching, 
a square lattice of holes with the diameter $\sim 120$\,nm and the period $a=200$\,nm
covering the $50 \times 120$\,$\mu$m$^2$ area, was created, see insets in Fig.\,\ref{fig:RT}.
The voltage probes V1-V2 were designed to fall within the nanopatterned 
domain of the film, while the probes V3-V4 were placed within its continuous
(non-perforated) section to measure thus the resistance of the original continuous film.  
The perforated section of the film confined between the probes V1-V2 contained 
50\,$\mu$m$\times$100\,$\mu$m$/(200$\,nm$)^2 = 125000$ elemental units.  
To increase the sample sheet resistance, it was sequentially treated 
by an additional soft plasma etching  two times. 
We will be referring to the untreated sample as to the original one (or as to the state A),
discriminating between the non-perforated (i.e. reference) part, rA 
and the perforated, pA, section of the film.  
Sequential etching transforms the film into the state B and eventually 
into the state C, and we will be using the prefixes `r' and `p' as in the state A.
The room temperature resistances are 3.17\,k$\Omega$ for rA,  3.75\,k$\Omega$ for rB, 
and 4.76\,k$\Omega$ for rC and 13.87\,k$\Omega$ for pA, 14.45\,k$\Omega$ for pB, 
and 16.26\,k$\Omega$ for pC.
The ratios of the room temperature resistances of the perforated 
and continuous samples for each respective state is about 3-4, reflecting 
the approximately $3$ times reduction of the effective cross-section 
of the sample upon perforation.
The chosen fabrication procedure allowed studying the evolution 
of electronic transport properties in a most controllable way avoiding introduction 
of additional geometric parameters that could vary from sample to sample. 
The temperature $T$- and the magnetic field $B$-dependences of the resistance were
measured using the standard four-probe dc and low frequency ac techniques.
The currents were sufficiently small to ensure the linear
response regime as was verified by direct measurements 
of the current-voltage characteristics $I$-$V$.   
The magnetic field was applied perpendicular to the film surface.

We start discussing our results with the zero magnetic field data. 
Shown in the Fig.\,\ref{fig:RT} are temperature dependences of the resistance 
per square (to obtain it the measured resistance was divided by two in accordance 
with the sample aspect ratio) for all three states of the perforated films 
[panels (a) and (c)] and for the corresponding reference films [panels (b) and (d)]. 
Both sets demonstrate the superconductor - insulator transitions (SIT).
To emphasize on the features of the temperature behaviour of the resistance,
we present resistance in both, linear and logarithmic scales, 
and the temperature in logarithmic sale since it spans over three orders of magnitude.
Resistances of all the films grow upon decreasing 
the temperature from the room temperature down, with all
the superconducting samples showing a nonmonotonic $R(T)$ behaviour with  
the pronounced maximum preceding the superconducting transition.  
So, the maximal resistance of the as-prepared reference superconducting film rA, 
is $R_{\mathrm{max}}=5.67$\,k$\Omega$.
The resistance of the film rB achieves $R_{\mathrm{max}}=10.3$\,k$\Omega$, 
and for the nanostructured film pA the maximal resistance value becomes 
$R_{\mathrm{max}}=39.4$\,k$\Omega$ at $T=1.64$\,K.
On the way down to $T_{\mathrm c}=1.03$\,K of the rA film 
the resistance of the pA sample shows a noticeable decrease to the value 
of $R=33.7$\,k$\Omega$, which is just about of the resistance per square of the rA film.  
Upon further cooling the pA sample exhibits the drop in the resistance of
about two orders of magnitude.
Nevertheless, the pA sample does not transit into a global phase coherent superconducting state,
remaining in the resistive state even at lowest temperatures. 
Deferring the discussion of this feature till after the presentation of the magnetoresistance data,
we stress here  a striking behaviour of the B-state.  
While the reference sample rB falls to a superconducting state 
(although it occurs at  $T_c=0.43$\,K, which is decreased as compared to that of rA), 
the nanostructured part pB appears at the insulating state of the SIT.
In the state C both, rC and pC are insulating.

\begin{figure}[tbh]
\includegraphics[width=1.0\columnwidth]{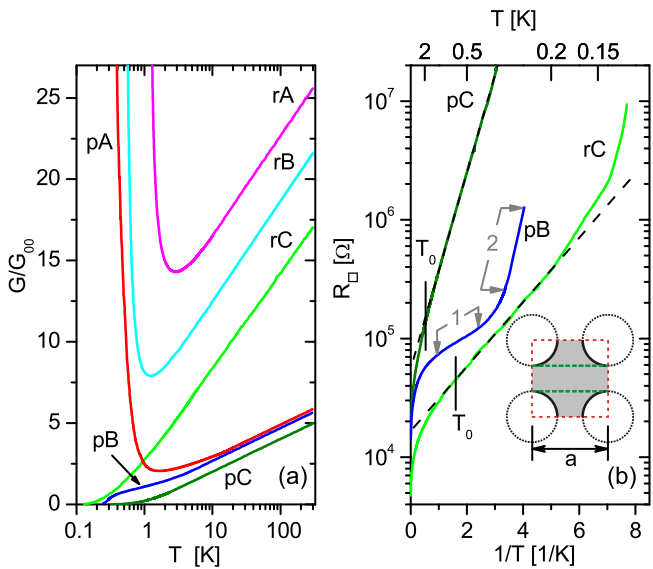}
\caption{\label{fig:Analysis}
(a) Conductance $G/G_{00}=2\pi^2 \hbar/(e^2R_\square)$ as function of temperature. 
Panel\,(b): $R_\square$ versus $1/T$ dependences for the non-superconducting samples.
Both rC and pC show Arrhenius behaviors, the dashed line depicts 
the fit to Eq.\,(1)  yielding characteristic temperatures 
$T_0=0.63$\,K and 1.9\,K, and $R_0=16.6$\,k$\Omega$ and 
55.6\,k$\Omega$ for samples rC and pC, respectively.
The line bar marks these $T_0$'s emphasizing that $T_0$'s 
were determined in the domain $T<T_0$ as proper.
At lowest temperatures $\log R(T)$ turns upwards
departing from the Arrhenius- for \textit{hyperactivated} regime~\cite{TiN_HA}.
The behavior of sample pB can not be viewed as
the Arrhenius one: Linearisation in the interval 1, going up from 0.4\,K  
results in $T_0=0.35$\,K, and the Arrhenius law cannot be used for $T>T_0$, 
while the linearisation in the interval 2 although giving acceptable 
$T_0=2.21$\,K, results in an unreasonably low $R_0=0.15$\,k$\Omega$.
The inset shows a geometry of an elemental unit of an array, 
illustrating the reduction of the cross section of the effective
conducting channel.}
\end{figure}

We now turn to details of the electronic transport properties.   
Upon cooling down to 10\,K all samples exhibit
logarithmic temperature dependence of the conductance
(see Fig.\,\ref{fig:Analysis}a), which is well described by the formula
$G(T)/G_{00}=A\ln(k_BT\tau/\hbar)$, where $G(T)=1/R_\square(T)$
is the conductance, $G_{00}=e^2/(2\pi^2\hbar)$.
This behavior is in accord with the theory of quantum corrections
for quasi-two-dimensional disordered systems and can be attributed
to weak localization and repulsive electron-electron interaction corrections~\cite{AAreview}.
Similar behaviour was observed before in critically disordered TiN 
films~\cite{STM_TiN,TiN_HA} and in Bi films~\cite{Goldman1993,Valles1999}.
Notably, the high-temperature slope is identical for all three samples within each sets 
(with $A=2.6\pm 0.1$ for the reference and $A=0.85\pm 0.05$ for perforated films),
irrespectively to their low-temperature either superconducting or insulating behaviors.
The ratio of the factors $A$ in the reference and perforated parts of the film 
is close to 3, as expected, due to the geometric reduction 
of the effective cross sections of the conducting channels 
in nanopatterned structures as compared to those in the reference films.
This evidences that patterning films does not introduce additional microscopic disorder.  
Thus disorder remains the same in perforated and continuous films for each state, A, B, and C.  

To analyse the behavior of the non-superconducting samples
we replot $R(T)$ as function of $1/T$ in Fig.\,\ref{fig:Analysis}b.
At low temperatures it is well fitted by the Arrhenius formula for both,
the reference and the nanopatterned films in the state C, 
evidencing that these samples are indeed insulators.
The dashed lines correspond to
\begin{equation}
R =R_0 \exp(T_0/T)\,,
\label{main}
\end{equation}
with $T_0=0.63$\,K and 1.9\,K being the characteristic temperatures,
and $R_0=16.6$\,k$\Omega$ and 55.6\,k$\Omega$ for samples rC and pC, respectively.
That $R_0$ of the pC is 3.35 times larger than that of the reference film, 
paralleled by the ratio of the respective resistances 
at room temperatures that equals to 3.4, gives more support to observation 
that patterning does not change the degree of microscopic disorder.

Speaking of the characteristic temperatures, one would have expected 
that had $T_0$ been formed over the microscopic scale, 
it would have remained the same in both reference and perforated films.  
Instead, one sees a noticeable increase in $T_0$ in a perforated sample.
This suggests that $T_0$ is built on the macroscopic spatial scales, 
not less then 200\,nm and is influenced by changes in geometric characteristics 
and the connectivity introduced by patterning.
Such a behaviour becomes clear, once one recalls that the characteristic energy 
of an insulating state of a two-dimensional Josephson junction array (JJA)
is $k_{\mathrm{B}}T_0=\Delta_{\mathrm{c}}=E_{\mathrm{c}}\ln (L/b)$, 
where $E_{\mathrm c}$ is the charging energy of a single superconducting island, 
$L$ is the smaller quantity out of either the electrostatic screening length 
of JJA or its linear dimension, and $b$ is the size of the elemental cell 
of the array~\cite{FVB,VinNature}.   
Maintaining that the Cooper-pair-insulating film comprises a self-induced texture 
of weakly coupled superconducting islands, one would have expected here the same behaviour
(the size-dependence of the characteristic energy was observed 
in the identically prepared TiN films~\cite{Kalok} and InO films~\cite{Ovadyahu}).
Then the upturn in the $\log R(T)$ vs. $1/T$ dependence for the rC sample, 
as well as for the perforated sample pB, occurs at the temperature 
of the charge Berezinskii-Kosterlitz-Thouless transition (CBKT) 
$k_{\mathrm B}T_{\mathrm{CBKT}}\simeq E_{\mathrm c}$; 
for the rC sample $T_{\mathrm{CBKT}}\approx 0.14$\,K. 
(In artificially manufactured 2D JJA the upturn from the 
behaviour described by the Eq.\,(1) due to CBKT
was observed in Refs.~\cite{Kanda,Yamaguchi}.)  
This yields $\ln(L/b)=\Delta_{\mathrm c}/E_{\mathrm c}=4.5$. 
In Ref.~\cite{FVB} the quantity $b \approx 40$\,nm was found for critically disordered  
TiN films in the vicinity of the SIT,
corresponding to $b\approx 4\xi$.
We, thus, obtain $L\approx 3.4\,\mu$m, and,  
since this value is apparently less then the size of the sample 
of 50\,$\mu$m (but at the same time well exceeds 200\,nm), 
$L$  represents the electrostatic screening length. 
To crudely estimate the effect of perforation we assume 
the originally square Josephson network with the lattice 
constant $\approx 40$\,nm and notice that perforation transforms 
this network into the new one with the 'bond' length equal to 200\,nm, 
each bond comprising a series of five junctions.   
This straightforwardly gives an estimate for new characteristic energy 
${\tilde\Delta}_{\mathrm{c}}=5\Delta_{\mathrm{c}}-5E_\mathrm{c}\ln 5=2$\,K, 
which nicely agrees with the experimental finding of 1.9\,K 
for the characteristic energy for the sample pC.  
The increase in the characteristic energy upon removing the fraction of the junctions 
of the 2D JJA was found in~\cite{Yamaguchi}.

\begin{figure}[t]
\includegraphics[width=1.0\columnwidth]{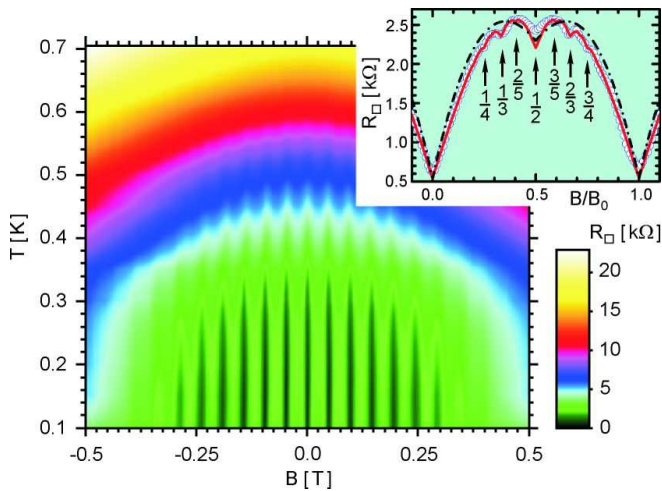}
\caption{\label{fig:Osc} 
The two-dimensional colour map of the resistance 
 in the temperature - magnetic field plane for the perforated sample pA. 
The colour scale quantifying magnitudes of the resistance is given at the lower right corner. 
The inset shows the experimental data of the magnetoresistance (open circles) vs. 
reduced magnetic field $B/B_0$, where $B_0= \Phi_0/a^2$, at temperature $0.11$\,K.
In addition to the fundamental dips at $B_n= B_0$, where $n$ is an integer, the
secondary dips at $B/B_0=1/4,\,1/3,\,2/5,\,1/2,\,3/5,\,2/3,$ and $3/4$ are observed.
Dash-dotted line corresponds to the periodic dependence of the fractional 
reduction in average Josephson coupling energy on the magnetic 
flux per unit cell derived in the nearest-neighbour contour
approximation of Ref.~\cite{Tinkham83}.  
The solid line passes through the theoretical values calculated 
for rational $B/B_0=p/q$, with $q\leq 256$,
on the superconducting network~Eq.\,(2) accounting 
for multi-contour phase synchronization and thus exhibiting 
higher order commensurability effects.
}
\end{figure}

Applying magnetic field to nanopatterned films brings in competing energy- 
and spatial scales and thus commensurability effects which manifest themselves 
through the oscillations in thermodynamics and transport properties.  
Shown in Fig.\,\ref{fig:Osc} are the magnetoresistance data for the pA sample 
at relatively weak magnetic field, over the range  $\pm 0.5$\,T,
which is much less than the upper critical field of the reference film 
(the reference film is close by its parameters to the low resistive sample 
of Ref.\,\cite{TiNPhysB2005}) $B_{\mathrm{c2}}=2.8$\,T.  
There are eight pronounced magnetoresistance (MR) oscillations at low temperatures 
(for each polarity of the field). 
The main period is $B_0=\Phi_0/a^2$, where $\Phi _0 =\pi \hbar /e$ 
is the superconducting flux quantum, corresponding one flux quantum per unit cell. 
Magnification of the resistive curve (see inset in Fig.\,\ref{fig:Osc}) 
reveals a fine structure reflecting 
collective behaviour of the multiconnected superconducting film 
differing it from the behavior of a single superconducting loop.
One distinguishes additional well-defined dips at 
$B/B_0=1/4,\,1/3,\,2/5,\,1/2,\,3/5,\,2/3,$ and $3/4$.
The previous observations reported the $B/B_0=1/2$ feature 
in square Josephson junction arrays~\cite{Webb83,Mooji1987,Mooji1992,Mooji1996}, 
proximity-effect junction arrays~\cite{Tinkham83,Kimhi84,Forrester88,Lobb1990}, 
perforated films and superconducting wire networks~\cite{Rammal1984,Moshchalkov2002,PhysBPtSi2003,PhysBPtSi2006}.  
The ``full" set of dips was observed in the MR of square JJA~\cite{Mooji1992,Mooji1996}
and in proximity-effect junction arrays~\cite{Lobb1990} 
and in the critical temperature variation in square 
superconducting wire network~\cite{Rammal1984}.

A description of the observed modulated MR is based 
on the solution of the Ginzburg-Landau equation on the superconducting network 
in the presence of the magnetic field~\cite{Rammal1984}.
The wave function $\psi$ for the square superconducting network follows the so-called
Harper equation~\cite{Harper1955}:
    \begin{equation}\label{eq:Harper}
      \psi_{n+1}+\psi_{n-1}+2\cos(2\pi\Phi n+\alpha)\psi_n=\epsilon \psi_n\,.
    \end{equation}
Here magnetic field $B$ appears through the flux per plaquette 
$Ba^2/\Phi_0\equiv\Phi/\Phi_0= p/q$, where $p$ and $q$ are relative prime numbers.  
The matrix, corresponding to this equation has $q\times q$ dimensions.  
The fields at which resistivity exhibits the dips are defined by the boundaries 
of the energy spectrum as function of the magnetic field,
i.e. one has to find $E(\Phi)=\max_\alpha[\epsilon(\Phi,\alpha)]$.  
To this end we decompose the matrix discriminant into the polinomial 
of $\varepsilon$ independent on $\alpha$ 
and the offset function independent on $\varepsilon$~\cite{Kreft93}. 
The change in the magnetoresistance is calculated as 
$\Delta R(B)=A\arccos^2[E(\Phi)/4]$, where $A$ is a numerical coefficient,
which does not depend on the magnetic field. 
The resulting $\Delta R(B)$ behaviour is shown by the solid line 
in the  Fig.\,\ref{fig:Osc} and demonstrates an
excellent agreement with the data at low temperatures.
The half-flux dip results mainly from the contribution from
the currents in the adjacent loops as it was shown in Ref.\,\cite{Tinkham83}. 
As temperature increases the fine structure is smeared out, 
while the oscillations with the main period $B_0$ persist
till $T\simeq0.7$\,K, where they are easily resolved by taking $dR/dB$.  

The most intriguing aspect of the observed MR oscillations is an extremely 
wide temperature region of their presence. 
The MR oscillations in the perforated films and/or superconducting wire networks (SWN), 
were usually found, if measured in the linear response regime, in the close proximity to $T_c$ 
in the region 
$\Delta T\lesssim 0.02 T_c$~\cite{Fiory1978,Rammal1984,Hoffmann2000,Moshchalkov2002,Kwok2007,NbN,KwokAPL2010}.
More extended temperature regions of the MR oscillations were reported in 
Refs.\,\cite{PhysBPtSi2003,PhysBPtSi2006,Hirata2005,Valles_PRL,Mironov2010,SochnikovNN,SochnikovPRB}.
Juxtaposing the data obtained for various systems and inspecting the systems'
geometric characteristics we observe that as a rule the low-temperature boundary 
for the MR oscillations to appear corresponds to the temperature 
at which the ratio $w/\xi_{\mathrm{d}}(T)\lesssim 5$,
where $w$ is the width of the superconducting constriction and
$\xi_{\mathrm d}(T)=\xi_{\mathrm d}(0)/\sqrt{1-T/T_{\mathrm{c}}}$.
This brings to the mind the Likharev's result~\cite{Likharev1979} 
that the weak link cannot accommodate an Abrikosov vortex if $w<w_{\mathrm{c}}$ 
and transforms into a Josephson junction, where the critical width 
$w_{\mathrm{c}}(L)$ was evaluated as being equal to $4.41\xi(T)$ for the long, 
$L\gg w$, link ($L$ is the link length) near $T_{\mathrm{c}}$.  
For the square link, $L=w$, $w_{\mathrm{c}}\approx 5\xi$ and can become 
well larger for the short weak links.
In order to gain the insight into the meaning of the boundary separating 
the dissipative behaviour governed by the vortex motion from the resistive state 
due to phase slips, let us employ the approach used in~\cite{Melnikov2006} 
to evaluate the condition of clustering 
of vortex cores.  
Namely, note that either in the presence of the magnetic field or upon passing the current, 
superconductivity near the edges of the constriction is suppressed 
and the Andreev states separated by minigaps 
$\simeq\Delta/(k_{\scriptscriptstyle{\mathrm{F}}}\xi)$ should form. 
Due to overlap of the wave functions localized at the opposite edges 
these states broaden as $\Delta\exp(-w/\xi)$. 
At the width $w_0\simeq\xi\ln(k_{\scriptscriptstyle{\mathrm{F}}}\xi)$, 
where the quasiparticle levels broadening 
becomes of the order of the level separation,  the constriction turns 
metallic and start to behave as a proximity 
effect generated Josephson junction.
The estimate for $w_0$ (at the plausible values of $k_{\scriptscriptstyle{\mathrm{F}}}$ 
and $\xi$) is in a reasonable 
agreement with the all available experimental data and gives the ratio $w_0/\xi(T)\simeq 4\div 8$.
Note that the above qualitative consideration does not use closeness to $T_{\mathrm{c}}$ and
implies that the concept of the critical size 
of the superconducting weak link below which it turns into 
a Josephson junction can be extended to low temperatures.  

The next observation is that in order to exhibit oscillations in the magntoresistance, 
the SWN should be in a resistive state. 
Thus the second condition determining the `range of observability' of the oscillations 
is that the temperature should be higher than the temperature of the vortex BKT transition 
in the SWN, $T_{\scriptscriptstyle{\mathrm{VBKT}}}$.  
Using the magnitude of the critical current
observed in our experiment at $T=0.1$\,K evaluated from the position of the maximum 
of the $dV/dI$ vs. $I$ curve~\cite{Lobb1990},
$I_{c}=0.17\,\mu$A, giving $I_{c(i)}=0.68$\,nA per one constriction,
one obtains the Josephson coupling energy as 
$E_{\scriptscriptstyle{\mathrm{J}}}/k_{\scriptscriptstyle{\mathrm{B}}}=0.016$\,K.  
Therefore, indeed, even the lowest temperature of our experiment (0.1\,K),
where the MR oscillations are clearly detectable, lie in the interval 
$T>T_{\scriptscriptstyle{\mathrm{VBKT}}}=\pi E_{\scriptscriptstyle{\mathrm{J}}}/
(2k_{\scriptscriptstyle{\mathrm{B}}})$~\cite{Lobb1983}.
For this reason the $R_\square(T)$ dependence for the pA sample remains 
finite at all the experimental temperatures.

\begin{figure}[b]
\includegraphics[width=1.0\columnwidth]{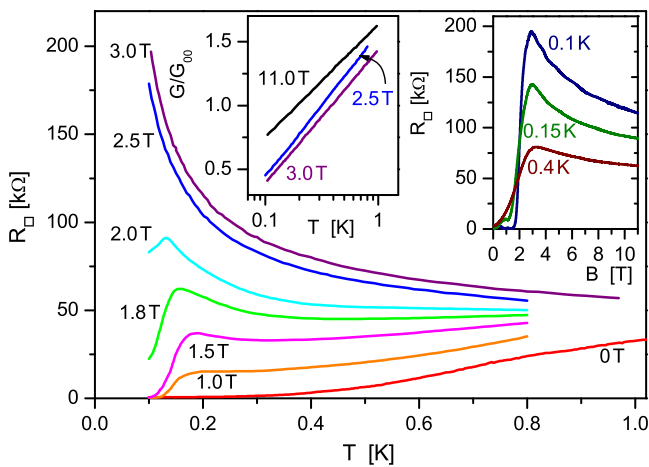}
\caption{\label{fig:RBhigh} 
The resistance per square versus temperature plots in the magnetic fields 
$B=0$, 1.0, 1.5, 1.8, 2.0, 2.5, 3.0\,T  
for the perforated sample pA. 
Left inset: The conductance $G/G_{00}=2\pi^2 \hbar/(e^2R_\square)$ 
as a function of the temperature 
in a logarithmic scale in the magnetic fields $B=2.5$, 3.0, and 11.0\,T.
Right inset: Magnetoresistance isotherms at temperatures
$T=0.1$, 0.15, and 0.4\,K.  
The positions of maxima are $B_{\mathrm{max}}=2.9$, $3.0$, and 3.3\,T,
correspondingly.
}
\end{figure}

Finally, we briefly discuss the MR at large fields beyond 
the range of oscillations for the pA sample.
At fields slightly larger than the critical field $B_{\mathrm{c2}}=2.8$\,T 
of the reference film~\cite{TiNPhysB2005}, we observe a huge peak of the MR 
at some $B=B_{\mathrm{max}}$, followed by the appreciable negative MR, 
as shown at the right inset in Fig.\,\ref{fig:RBhigh}.
The value of $B_{\mathrm{max}}$ shifts towards larger fields upon increase in $T$.  
The similar shifts of the maxima and the subsequent negative MR 
are observed in the reference film (see Fig.\,1c of Ref.\,\cite{TiNPhysB2005}) 
and are well described by the quantum corrections 
to the conductivity due to superconducting fluctuations~\cite{GalLar}.  
The main panel of Fig.\,\ref{fig:RBhigh} shows the evolution 
of the $R_{\square}(T)$ dependences at different magnetic fields. 
At zero field there is a high temperature, $T=1.64$\,K, maximum (Fig.\,\ref{fig:RT}).
At higher fields, but not too far beyond $B\approx 2$\,T, a second, 
low temperature maximum appears.
At yet higher fields, as $B$ approaches and further exceeds $B_{\mathrm{c2}}$ 
of the reference film, the $R_{\square}(T)$-dependences become monotonically 
decreasing with the temperature increase.
Notably, in spite of the fact that the MR exhibits huge peak, up to 200\,k$\Omega$ 
per square at 0.1\,K, it can be hardly viewed as the manifestation of the magnetic-field-induced 
superconductor-insulator transition.
Indeed, the conductance $G(T)=1/R_{\square}(T)$ at fields around and above the peak, 
fits perfectly to the logarithmic temperature dependence $G(T)/G_{00}=A\ln(k_BT\tau/\hbar)$,
see left inset in Fig.\,\ref{fig:RBhigh},
with $A$ monotonically decreasing with the growth of the magnetic field, $A=0.49$, 0.45, and 
0.38 at fields 2.5, 3.0, and 11.0\,T, respectively.  
This decrease is likely to reflect the suppression of the
contribution from the superconducting fluctuations to conductivity.  
Moreover, we observe that the value of $A=0.38$, being multiplied by the geometry factor of 3, 
becomes equal to 1.14, which is pretty close $A=1$ expected for
the contribution to the conductivity of disordered metals due to 
the repulsive electron-electron interaction~\cite{AAreview}.
We conclude that although this sample does not exactly exhibits the magnetic 
field driven SIT, it still demonstrates the significant amplification 
of the resistivity growth with the decreasing temperature
(the resistance increases by almost the factor of 3.5 
in the interval from 1 to 0.1\,K at $B=3$\,T, while in the reference film 
the corresponding factor is about 1.4),
and thus can be considered as being put at the threshold of the transition 
from the weak to strong localization by nanopatterning. 

In summary, we have shown that nanopatterning disordered superconducting films 
pushes the SIT to the lower degree of microscopic disorder opening the route to control 
the position of the SIT on the phase diagram.  
We have revealed a wide spectrum of phenomena related to
periodicity of the phase and the absolute value of the superconducting 
order parameter in a superconducting network.  

\begin{acknowledgments}
We are delighted to thank Boris Shapiro (Bar-Ilan University) and Alexander Mel'nikov for
useful discussions.
This research is supported by the Program ``Quantum Physics of Condensed Matter'' 
of the Russian Academy of Sciences, by the Russian Foundation for Basic Research 
(Grant No. 09-02-01205), and by
the U.S. Department of Energy Office of Science under the Contract
No. DE-AC02-06CH11357. 
\end{acknowledgments}


\begin{references}


\bibitem{Strongin70} 
M.~Strongin, R.\,S.~Thompson, O.\,F.~Kammerer, and J.\,E.~Crow, 
Phys. Rev. B \textbf{1}, 1078 (1970).

\bibitem{Haviland} 
D.\,B.~Haviland, Y.~Liu, and A.\,M.~Goldman,
Phys. Rev. Lett. \textbf{62}, 2180 (1989).

\bibitem{Goldman1993}
Y.~Liu, D.\,B.~Haviland, B.~Nease, and A.\,M.~Goldman,
Phys. Rev. B \textbf{47}, 5931 (1993).

\bibitem{GoldmanReview} 
A.\,M.~Goldman and N.~Markovi\'c,
Phys. Today \textbf{51}(11), 39 (1998).

\bibitem{BeWu} 
E.~Bielejec, J.~Ruan, and W.~Wu,
Phys. Rev. Lett. \textbf{87}, 036801 (2001).

\bibitem{TiNSanquer}
N.~Hadacek, M.~Sanquer, and  J.-C.~Vill\'{e}gier, 
Phys. Rev. B \textbf{69}, 024505 (2004).

\bibitem{TaYoon} 
Y.~Qin, C.\,L.~Vicente, and J.~Yoon,
Phys. Rev. B \textbf{73}, 100505(R) (2006).

\bibitem{Maekawa} 
S.~Maekawa and H.~Fukuyama,
J. Phys. Soc. Jpn. \textbf{51}, 1380 (1982).

\bibitem{Finkelstein} 
A.\,M.~Finkel'stein,
Sov. Phys. JETP Lett. \textbf{45}, 37 (1987);
Physica B \textbf{197}, 636 (1994).

\bibitem{TiNPhysB2005}
T.\,I.~Baturina, J.~Bentner, C.~Strunk, M.\,R.~Baklanov, A.~Satta,
Physica B \textbf{359}, 500 (2005).

\bibitem{QMTiNPRL}
T.\,I. Baturina, C. Strunk, M.\,R. Baklanov, and A. Satta,
Phys. Rev. Lett. \textbf{98}, 127003 (2007).

\bibitem{SITTiNPRL}
T.\,I. Baturina, A.\,Yu. Mironov, V.\,M. Vinokur,
M.\,R. Baklanov, and C. Strunk,
Phys. Rev. Lett. \textbf{99}, 257003 (2007).

\bibitem{QRCTiNPhysB}
T.\,I. Baturina, A.~Bilu\v{s}i\'{c}, A.\,Yu. Mironov, V.\,M.
Vinokur, M.\,R. Baklanov, and C. Strunk,
Physica C \textbf{468}, 316 (2008).

\bibitem{OpticsTiN}
F.~Pfuner, L.~Degiorgi, T.\,I.~Baturina, V.\,M.~Vinokur, and M.\,R.~Baklanov,
New Journal of Physics \textbf{11}, 113017 (2009).

\bibitem{STM_TiN}
B.\,Sac\'{e}p\'{e}, C.\,Chapelier, T.\,I.\,Baturina,
V.\,M.\,Vinokur, M.\,R.\,Baklanov, and M. Sanquer,
Phys. Rev. Lett. \textbf{101}, 157006 (2008); 
preprint arXiv:0906.1193.
                     
\bibitem{AAreview}
B.\,L.~Altshuler and A.\,G.~Aronov,
in: {\it Electron-Electron Interactions in Disordered Systems},
ed. by A.\,L.~Efros and M.~Pollak (North-Holland, Amsterdam, 1985).

\bibitem{TiN_HA}
T.\,I.~Baturina, A.\,Yu.~Mironov, V.\,M.~Vinokur, M.\,R.~Baklanov, and C.~Strunk,
JETP Lett. \textbf{88}, 752 (2008).

\bibitem{Valles1999}
J.\,A.~Chervenak and J.\,M.~Valles, Jr.,
Phys. Rev. B \textbf{59}, 11209 (1999).

\bibitem{FVB} 
M.\,V.~Fistul, V.\,M.~Vinokur, and T.\,I.~Baturina,
Phys. Rev. Lett. \textbf{100}, 086805 (2008).

\bibitem{VinNature}
V.\,M.~Vinokur, T.\,I.~Baturina, M.\,V.~Fistul,
A.\,Yu.~Mironov, M.\,R.~Baklanov, and C. Strunk, 
Nature \textbf{452}, 613 (2008).

\bibitem{Kalok} 
D.~Kalok, A.~Bilu\v{s}i\'{c}, T.\,I.~Baturina,  V.\,M.~Vinokur, and C.~Strunk,
to be published.

\bibitem{Ovadyahu} 
D.~Kowal and Z.~Ovadyahu,
Physica C \textbf{468}, 322 (2008).

\bibitem{Kanda} 
A. Kanda and S. Kobayashi,
J. Phys. Soc. Jpn. \textbf{64}, 19 (1995).

\bibitem{Yamaguchi} 
T. Yamaguchi, R. Yagi, S. Kobayashi, and Y. Ootuka,
J. Phys. Soc. Jpn. \textbf{67}, 729 (1998).

\bibitem{Webb83} 
R.\,A.~Webb, R.\,F.~Voss, G.~Grinstein, and P.\,M.~Horn,
Phys. Rev. Lett. \textbf{51}, 690 (1983).

\bibitem{Mooji1987} 
B.\,J.~van Wees, H.\,S.\,J.~van der Zant, and J.\,E.~Mooij,
Phys. Rev. B \textbf{35}, 7291 (1987). 

\bibitem{Mooji1992} 
H.\,S.\,J.~van der Zant, F.\,C.~Fritschy, W.\,J.~Elion, L.\,J.~Geerligs, and J.\,E.~Mooij,
Phys. Rev. Lett. \textbf{69}, 2971 (1992). 

\bibitem{Mooji1996} 
H.\,S.\,J.~van der Zant, W.\,J.~Elion, L.\,J.~Geerligs, and J.\,E.~Mooij,
Phys. Rev. B \textbf{54}, 10081 (1996). 

\bibitem{Tinkham83}  
M.~Tinkham, D.\,W.~Abraham, and C.\,J.~Lobb,
Phys. Rev. B \textbf{28}, 6578 {1983}.

\bibitem{Kimhi84} 
D.~Kimhi, F.~Leyvras, and D.~Ariosa,
Phys. Rev. B \textbf{29}, 1487 (1984).

\bibitem{Forrester88} 
M.\,G.~Forrester, Hu Jong Lee, M.~Tinkham, and C.\,J.~Lobb,
Phys. Rev. B \textbf{37}, 5966 (1988).

\bibitem{Lobb1990} 
S.\,P.~Benz, M.\,S.~Rzchowski, M.~Tinkham, and C.\,J.~Lobb,
Phys. Rev. B \textbf{42}, 6165 (1990).

\bibitem{Rammal1984} 
B.~Pannetier, J.~Chaussy, R.~Rammal, and J. C. Villegier,
Phys. Rev. Lett. \textbf{53}, 1845 (1984).

\bibitem{Moshchalkov2002}
L.~Van Look, B.\,Y.~Zhu, R.~Jonckheere, B.\,R.~Zhao, Z.\,X.~Zhao, and V.\,V.~Moshchalkov,
Phys. Rev. B \textbf{66}, 214511 (2002).

\bibitem{PhysBPtSi2003} 
T.\,I.~Baturina, D.\,W.~Horsell, D.\,R.~Islamov, I.\,V.~Drebushchak,
Yu.\,A.~Tsaplin, A.\,A.~Babenko, Z.\,D.~Kvon, A.\,K.~Savchenko, A.\,E.~Plotnikov,
Physica B \textbf{329}, 1496 (2003).

\bibitem{PhysBPtSi2006} 
T.\,I.~Baturina, Yu.\,A.~Tsaplin, A.\,E.~Plotnikov, M.\,R.~Baklanov,
Physica B \textbf{378}, 1058 (2006).

\bibitem{Harper1955}
P.\,G.~Harper, Proc. Phys. Soc. London A \textbf{68}, 874 (1955).

\bibitem{Kreft93}
C. Kreft, TU-Berlin, SFB 288, Preprint No. 89 (1993).

\bibitem{Fiory1978}
A.\,T.~Fiory, A.\,F.~Hebard, and S.~Somekh, 
Appl. Phys. Lett. \textbf{32}, 73 (1978).

\bibitem{Hoffmann2000}
A.~Hoffmann, P.~Prieto, and I. K. Schuller,
Phys. Rev. B \textbf{61}, 6958(R) (2000).

\bibitem{Kwok2007}
U.~Patel, Z.\,L.~Xiao, J. Hua, T. Xu, D. Rosenmann, V. Novosad, J. Pearson, U. Welp,
W.\,K.~Kwok, and G.\,W.~Crabtree,
Phys. Rev. B \textbf{76}, 020508(R) (2007).

\bibitem{KwokAPL2010} 
S. Avci, Z. L. Xiao, J. Hua, A. Imre, R. Divan, J. Pearson, U. Welp, W. K. Kwok, and G. W. Crabtree,
Appl. Phys. Lett. \textbf{97}, 042511 (2010).

\bibitem{NbN}
Ajay D. Thakur, Shuuichi Ooi, Subbaiah P. Chockalingam, John Jesudasan,
Pratap Raychaudhuri, and Kazuto Hirata,
Appl. Phys. Lett. \textbf{94}, 262501 (2009).


\bibitem{Valles_PRL}
H.\,Q.~Nguyen, S.\,M.~Hollen, M.\,D.~Stewart, Jr., J.~Shainline, 
Aijun Yin, J.\,M.~Xu, and J. M. Valles, Jr.,
Phys. Rev. Lett. \textbf{103}, 157001 (2009).


\bibitem{Hirata2005}
S. Ooi, T. Mochiku, S. Yu, E.\,S. Sadki, K. Hirata,
Physica C \textbf{426}, 113 (2005).

\bibitem{Mironov2010}
A.\,Yu.~Mironov, T.\,I.~Baturina, V.\,M.~Vinokur, S.\,V.~Postolova, 
P.\,N.~Kropotin, M.\,R.~Baklanov, D.\,A.~Nasimov, A.\,V.~Latyshev,
Physica C (2010), doi:10.1016/j.physc.2009.12.057.


\bibitem{SochnikovNN}
I. Sochnikov, A. Shaulov, Y. Yeshurun, G. Logvenov, and I. Bo\v{z}ovi\'{c},
Nature Nanotechnology \textbf{5}, 516 (2010).

\bibitem{SochnikovPRB}
I. Sochnikov, A. Shaulov, Y. Yeshurun, G. Logvenov, and I. Bo\v{z}ovi\'{c},
Phys. Rev. B \textbf{82}, 094513 (2010).


\bibitem{Likharev1979}
K.\,K.~Likharev, Rev. Mod. Phys. \textbf{51}, 101 (1979).

\bibitem{Melnikov2006}
A.\,S.~Mel'nikov and M.\,A.~Silaev,
JETP Lett. \textbf{83}, 578 (2006).
  
\bibitem{Lobb1983}
C.\,J.~Lobb, D.\,W.~Abraham, and M. Tinkham,
Phys. Rev. B \textbf{27}, 150 (1983).
  
\bibitem{GalLar}
V. M. Galitski  and A. I. Larkin,
Phys. Rev. B \textbf{63}, 174506 (2001).

\end{references}
\end{document}